\documentclass[prl,showpacs,superscriptaddress]{revtex4-1}
\usepackage[english]{babel}
\usepackage[utf8]{inputenc}
\usepackage{amsmath,amssymb,physics}
\usepackage{ragged2e}
\usepackage{graphicx}
\usepackage [T1] {fontenc}
\usepackage{color}
\usepackage{bm}
\usepackage{cases}

\newcommand{\sout}[1]{\unskip}

\newcommand{\bx}{{\bf x}}

\newcommand{\smeq}{ \! = \!}

\newcommand{\healing}{\nu}

\usepackage{hyperref}
\usepackage{amsfonts, vmargin}
\usepackage{tikz}
\usepackage{epic, eepic}
\usepackage[us]{datetime}
\usepackage{wrapfig}

\definecolor{BLACK}{named}{black}
\definecolor{GREEN}{named}{green}

\newcommand{\chkout}[1]{\unskip}

\begin{document}
\title{\sout{Potential representation of a Mean Field Game in the weak noise limit}
Universal behavior in non stationary Mean Field Games}

\author{Thibault Bonnemain}
\affiliation{LPTMS, CNRS, Univ.\ Paris-Sud, Universit\'e
  Paris-Saclay, 91405 Orsay, France}   
\affiliation{LPTM, CNRS, Univ.\ Cergy-Pontoise, 95302 Cergy-Pontoise, France}
\author{Thierry Gobron}
\affiliation{LPTM, CNRS, Univ.\ Cergy-Pontoise, 95302 Cergy-Pontoise, France}
\author{Denis Ullmo}
\affiliation{LPTMS, CNRS, Univ.\ Paris-Sud, Universit\'e Paris-Saclay,
  91405 Orsay, France}

\begin{abstract}
	Mean Field Games provide a powerful framework to analyze the dynamics of a large number of controlled objects in interaction. Though
	these models are much simpler than the underlying differential games they describe in some limit, their behavior
	is still far from being fully understood. 
	When the system is confined,  a notion of ``ergodic state'' has been introduced that characterizes most of
        the dynamics for long optimization times. Here we consider a
        class of models without such an ergodic state, and show the existence of a scaling solution that
        plays similar role. Its universality and scaling behavior can be inferred from a mapping to an electrostatic problem.
\end{abstract}

\date{\today}
\maketitle

Mean Field Games are a powerful framework introduced about a decade
ago by Lasry and Lions \cite{LasryLions2006-1} as an alternative
approach to differential game theory when the number of agents becomes
large. Their applications are numerous, ranging from finance
\cite{Lachapelle2015,lasry_mean_2007} and economy
\cite{GueantLasryLions2010,Achdou2014} to engineering sciences
\cite{WirelessNetwork,KizilkaleMalhame2015}, 
  and wherever one has to deal with optimization issues
for many coupled subsystems. On a quite general
basis, a Mean Field Game involves a set of $N$ players (or
agents) which are characterized by a continuous ``state variable''
${\bf X}^i \in \mathbb{R}^d$, $i=1 \dots N$, which, depending on the
context, may represent a physical position, the amount of resources
owned by a company, the house temperature in a network of controlled
heaters, etc..  These state variables evolve on a time interval
  $[0,T]$ according to some controlled dynamics, which we assume here
to be described by a linear, d-dimensional, Langevin equation,
$d {\bf X}^i_t = {\bf a}^i_t dt + \sigma d{\bf W}^i_t$, where each  component of ${\bf W}^i$ is an independent white noise of
variance one, $\sigma$ is a constant, and the ``control parameter'' is
the velocity ${\bf a}^i_t$. This control is adjusted in time by the
agent $i$ in order to minimize a cost functional which in the simplest case can be assumed of the form
\begin{eqnarray}\label{CostFunctional}
  c[{\bf a^i}]({\bf x}^i_t,t) & =    
  \mathbb{E} \left[ \int_{t}^T \left(\frac{\mu}{2}({\bf a}^i_t)^2 -
                             V[m_{\tau}]({\bf X}^i_\tau)\right)
                             d\tau \right.  \notag  \\ 
  &  + c_T({\bf X}^i_T) \biggr] \; . \label{eq:cost}
\end{eqnarray}

In \eqref{CostFunctional}, $V[m_{t}](\bx)$ is a functional of the empirical density
$m_t({\bf x}) = \frac{1}{N} \sum_j \delta ({\bf x} - {\bf X}^j(t))$,
 through which the agents' optimization problems are coupled. We shall assume $V[m_{\tau}](\bx)$
takes the simple form
\begin{equation}\label{potential}
V[m_{\tau}](\bx) = U_0(\bx) + g\;m_t(\bx) \; .
\end{equation}
For a very large number of players, like in a mean field theory, the
fluctuations of the empirical density are neglected and $m_t({\bf x})$
becomes a deterministic quantity governed by a Fokker Planck
equation. Furthermore, the optimization problems decouple and the
optimal value of the cost \eqref{CostFunctional} (the ``value function'') for the agent $i$
  becomes a function of its state variable
$\bx^i_t $ at time $t$,  solution of an Hamilton-Jacobi Bellman 
equation \cite{bertsekas2012}.  The
  resulting Mean Field Game is  defined as a pair
  of coupled equations,  a forward diffusion  for the
density $m(\bx,t)$ and a backward optimization for the
value function $u(\bx,t)$, which, in the simple case  we
consider here takes the form \cite{ULLMO20191}
\begin{numcases}{}
\partial_t m-\frac{1}{\mu}\nabla\left[m\nabla u\right]-\frac{\sigma^2}{2}\Delta m=0 & \label{FP}\\
\partial_t u-\frac{1}{2\mu}\left[\nabla
  u\right]^2+\frac{\sigma^2}{2}\Delta u=V[m] \; . \label{HJB}
\end{numcases}

The coupling between the two PDE's comes from two parts: in the
Fokker-Planck equation \eqref{FP}, the optimal velocity appears in the
drift term and here is proportional to the gradient of the value
function, ${\bf a} = -\frac{1}{\mu} \nabla u$; in the Hamilton-Jacobi
Bellman equation \eqref{HJB} the term $V[m]$ reflects the dependence
of the cost functional \eqref{CostFunctional} on the density. This
structure also induces rather atypical boundary conditions: the
(forward) Fokker-Planck equation is associated with an initial
condition $m(0,{\bf x})=m_0({\bf x})$ specifying the initial
distribution of agents, while the terminal cost in
\eqref{CostFunctional} imposes a final condition for the value
function, $u(T,{\bf x})=c_T({\bf x})$. This forward-backward structure
together with mixed initial-final boundary conditions leads to new
challenges when trying to characterize, either analytically or
numerically, solutions to this system of equations.

For a large class of settings, and in particular for repulsive
interactions, such a system, when confined, exhibits an ergodic
(stationary) state, independant on the boundary conditions, which can
be rigorously defined in the limit $T\to\infty$ as an hyperbolic fixed
point. The importance of this result, as proven by Cardialaguet et
al.\ \cite{Cardaliaguet2013} is twofold: for finite but long enough
optimization time $T$, the game will stay very close to this ergodic
(time independent) state except possibly in its initial and final
parts. Furthermore, the transient dynamics near $t=0$ and $t=T$
completely decouple one from the other, and the mixed boundary problem
simplifies accordingly: they describe how the initial and final
boundary conditions match with the ergodic state and are characterized
by two possibly different time scales, $\tau_0$ and $\tau_T$.

If however we consider a system of agents with repulsive interactions
in an unbounded state space, any initially localized configuration
will expand forever and no ergodic state will ever be reached. The
natural question we address here is whether some kind of limiting
regime still exists for such systems that would, at some level, play a
similar role as the ergodic state in confined systems.

\begin{figure}[hbt]
	\includegraphics[scale=0.3]{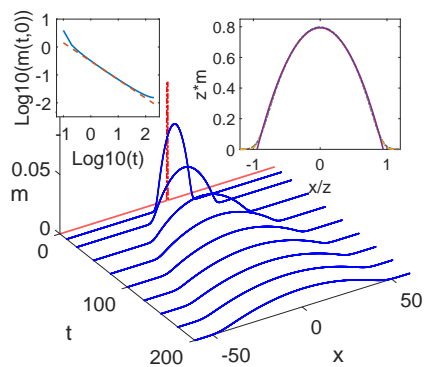}
	\caption{Time evolution of the density of agents. In this case
          $T=200$, $g=-2$, $\sigma=0.5$ and $\mu=1$. The initial
          distribution is a Gaussian of variance $0.1$, and the
          terminal cost is flat $c_T(x)=0$. The left inset shows the
          evolution of $m(t,0)=3/4z(t)$ with time (full line) compared
          to the $t^{2/3}$ scaling behavior (dashed). On the left
          inset are ploted an inverted parabola, and the density
          $m(x,t)$ rescaled by $z(t)$ for three different times.}
	\label{fig:m}
\end{figure}

A first indication that this is indeed the case is brought out by
numerical simulations. In Fig.~\ref{fig:m}, we report
   the numerical evolution of a one-dimensional
  system with a narrow initial distribution of
agents with repulsive interactions,  in the absence of  external
potential ($V[m_t](x)=gm(t,x)$, $g<0$). 
What we observe there is that except for transient times near $t=0$
and near $t=T$, the density assumes an almost perfect ``inverted
parabola form'' at all times
\begin{equation}
\label{eq:paransatz}
m(t,x)=\left\{
\begin{aligned}
&\frac{3}{4} \frac{(z(t)^2-x^2)}{z(t)^3} & \text{if $x \le z(t)$} \\
&0 & \text{ otherwise}
\end{aligned}
\right. \; ,
\end{equation}
where the prefactor derives from the normalization,
and the time dependent scaling factor
 $z(t)$  grows in time as a power law, 
\begin{equation} \label{eq:scaling}
z(t) \sim t^{2/3} \; .
\end{equation}
These features appear for sufficiently long optimization time $T$, and
are essentially independent of the initial and final boundary
conditions provided the initial distribution has a bounded extension
and the final cost is close to zero everywhere.

Another necessary condition for this behavior to appear is the
smallness of the healing length $\healing={\mu\sigma^2}/{|g|}$ with
respect to other lengthscales, but it will be eventually fulfilled
once the density distribution
  gets sufficiently extended (and at all times thereafter), possibly
    inducing a time shift in \eqref{eq:scaling}.

 What those numerical results tell us is that in this
 configuration, the notion of ergodic state has been replaced by the
 next best thing, namely a universal scaling solution.  The goal of
 this paper is to understand this puzzling result and in particular the $\frac{2}{3}$ scaling exponent. 
 
 We now introduce the few formal transformations which allow to show that this result is 
 rather natural and intuitive, once the problem is cast in its proper language. In turn, we will also gain a
 better understanding of this regime

\paragraph*{Hydrodynamic representation}

The main idea underlying these transformations comes from the deep
link between Mean Field Games and the nonlinear Schrödinger equation,
discussed at lengths in \cite{ULLMO20191}, which allows for the
use of several techniques developed to study Bose-Einstein
condensates. One of these techniques, the Madelung substitution, is
particularly well suited to deal with the small $\healing$ regime
\cite{pethick_smith_2008}. It consists in defining a velocity field
$v(t,x)$
\begin{equation}\label{v}
v(x,t) = -\frac{\nabla
  u}{\mu}-\frac{\sigma^2\nabla m}{2m}\; ,
\end{equation}
which drives the evolution of the density $m$ as a
simple transport/continuity equation
\begin{equation} \label{mhydro}
\partial_t m + \nabla(mv) = 0 \; .
\end{equation}
The evolution for the velocity field $v$ derives from the HJB equation \eqref{HJB}
\begin{equation}\label{vhydro}
\partial_t v+\nabla\left[\frac{\sigma^4}{2\sqrt m}\Delta\sqrt m+\frac{v^2}{2}+\frac{g}{\mu}m
\right]=0 \; ,
\end{equation}
and involves a $O(\sigma^4)$ term.  As in the context of cold atoms,
 this term can be neglected as long as
the characteristic length of the system is large in front of the
healing length $\healing={\mu\sigma^2}/{|g|}$
\cite{pethick_smith_2008}, leading to what is referred to as the
Thomas Fermi approximation.  This weak noise limit is also one of the
  requirements for the appearance of the parabolic behavior described
  in Figure \ref{fig:m}. In this approximation the equation read
\begin{equation} \label{hydroMFG}
\left\{
\begin{aligned}
&\partial_t m + \nabla(mv) = 0 \\
&\partial_t v+\nabla\left[\frac{v^2}{2}+\frac{g}{\mu}m\right] = 0 \\
\end{aligned}
\right. \; .
\end{equation}
These equations are formally very close to those studied for instance
in the field of cold atoms \cite{Dalfovo1999Theory}, the main
differences being the (negative) sign of $g$,
which makes the system elliptic rather than hyperbolic, and the nature
of the boundary conditions. Within this 
  approximation, it can easily be verified that
the ansatz Eq.~\eqref{eq:paransatz} with
\begin{equation}
\label{zansatz}
z(t)=3 \left(\frac{|g|}{4 \mu} \right)^{1/3} \, t^{2/3} \; , 
\end{equation}
is a particular solution of the Thomas-Fermi-like equations 
Eq.~\eqref{hydroMFG}. The real mystery is therefore not that such a
solution does exist, but to understand {\em why it is universal} in the
large optimization time limit.

\paragraph*{Riemann invariant and hodograph transform}
To answer this question, we shall turn to an approach developed in the
context of non-linear waves \cite{kamchatnov2000nonlinear}, which
relies on the notions of Riemann invariants and hodograph
transform. Riemann's method can be considered an extension of the
method of characteristics. It amounts to finding curves
(characteristics) on which some quantities (Riemann invariants) are
conserved. Here, one can show that there exists a pair
of Riemann invariants,
$(\lambda_+(x,t),\lambda_-(x,t))$, namely
$\lambda_\pm=v \pm 2 i \sqrt{{|g|m}/{\mu}}$,  so
that \eqref{hydroMFG}  reads
\begin{equation}\label{Rinv}
\left\{
\begin{aligned}
&\partial_t \lambda_++\left(\frac{3}{4}\lambda_++\frac{1}{4}\lambda_-\right)\partial_x\lambda_+=0\\
&\partial_t \lambda_-+\left(\frac{1}{4}\lambda_++\frac{3}{4}\lambda_-\right)\partial_x\lambda_-=0
\end{aligned}
\right. \; .
\end{equation}
Though characteristics do not exist in this context (they are curves
in the complex plane $\mathbb C^2$), this change of variables still
allows us to linearize these equations using an hodograph
transformation \cite{kamchatnov2000nonlinear}.  
Taking the pair $(\lambda_+,\lambda_-)$ as independant variables, we
express $x$ and $t$ as functions of them, so that the system
\eqref{Rinv} transforms into a linear one: 
\begin{equation}\label{hodoMFG}
\left\{
\begin{aligned}
&\partial_{\lambda_-}x-\beta_+\partial_{\lambda_-}t=0\\
&\partial_{\lambda_+}x-\beta_-\partial_{\lambda_+}t=0
\end{aligned}
\right. \; ,
\end{equation}
where $\beta_\pm=\left(\frac{3}{4}\lambda_\pm+\frac{1}{4}\lambda_\mp\right)$.
This system can be readily integrated once as
\begin{equation}\label{hodint}
\left\{
\begin{aligned}
&x-\beta_+t=\omega_+\\
&x-\beta_-t=\omega_-
\end{aligned}
\right. \; ,
\end{equation}
with $\omega_\pm$ solution of
\begin{equation}
\partial_{\lambda_\pm}\omega_{\mp}=-(\partial_{\lambda_\pm}\beta_{\mp})t
= -\frac{1}{4} t  \; .
\end{equation}
Thus the functions $\omega_\pm$ can be expressed 
as derivatives of a potential $\omega_\pm=\partial_{\lambda_\pm}\chi$, where $\chi(\lambda_+,\lambda_-)$
is solution of an Euler-Poisson-Darboux equation: 
\begin{equation}\label{Euler-Poisson}
\partial_{\lambda_+ \lambda_-} \chi-\frac{1}{2(\lambda_+-\lambda_-)}
(\partial_{\lambda_+}\chi-\partial_{\lambda_-}\chi) = 0 \; . 
\end{equation}
The main difference with the traditional treatment of NLS that we have
closely followed until now is that here the Riemann invariants
$(\lambda_+,\lambda_-)$ are complex conjugates.  In terms of its real
and imaginary parts, $\xi = \frac{1}{2} (\lambda_+ +\lambda_- ) =   v$ and
$\eta = \frac{1}{2i} (\lambda_+ -\lambda_- ) = \; 2\sqrt{{|g|m}/{\mu}}$, equation \eqref{Euler-Poisson} becomes 
\begin{equation} \label{lapcyl}
\partial_{\xi \xi}\chi+\partial_{\eta \eta} \chi + \frac{1}{\eta}\partial_{\eta}\chi=0 \; ,
\end{equation}
which is the Laplace equation in cylindrical
coordinates (with no angular dependence) with $\eta$ and $\xi$
as radial and axial coordinates, respectively.  Equations~\eqref{hodint}
now read
\begin{equation}\label{eq:E}
\left\{
\begin{aligned}
&\sout{\partial_{\eta}\chi=}\eta t= -E_\eta  \\
&\sout{\partial_{\xi}\chi=}2(x-\xi t)= - E_\xi
\end{aligned}
\right. \; ,
\end{equation}
with $E_\eta$ and $E_\xi$ the radial and axial components of the
electric field, ${\bf E} = - \nabla \chi$.  Note that even if
\eqref{lapcyl} is originally a two-dimensional problem, a clear
connection with electrostatics emerges when considering it a three
dimensional one with axial symmetry.

\paragraph*{Potential representation} 
Through the hodograph transform we have shown that for any potential
$\chi$, solution of the Laplace equation \eqref{lapcyl}, there is a
solution to the Thomas Fermi equation \eqref{hydroMFG}
  provided that the
  relations \eqref{eq:E} between $x$, $t$ and the electric field
  $\bf E$ hold.  The linear Laplace equation (and the related
  electrostatic problem) is clearly significantly simpler than the
original non-linear hydrodynamic equations.  The price to pay for that
simplification is that taking into account the
boundary conditions becomes highly non trivial since the locus of the
curves $t(\xi,\eta) = 0$ or $t(\xi,\eta) = T$ on which these
conditions are expressed actually depend on the particular potential
$\chi(\xi,\eta)$ considered.

Since the dynamics we are interested in is related to the spreading of
the density of agents, the time evolution is associated with a
contraction towards the origin in the plane $(\xi,\eta)$ and the equal
time curves, $t(\xi,\eta) = \text{const}$, are nested together, larger
times closer to the origin.  If we consider $\chi$ as generated by a
distribution of charge $\rho(\xi,\eta)$, Eq.~\eqref{lapcyl} implies
that there is no charges between the curves $t(\xi,\eta) = 0$ and
$t(\xi,\eta) = T$ but $\rho(\xi,\eta)$ can be non-zero either near the
origin (inside the curve $t(\xi,\eta) = T$) or at large distance
(outside the curve $t(\xi,\eta) = 0$).  If the optimization time is
long enough we can assume that there exists a large time inteval
$[\tilde t_{\rm min}, \tilde t_{\rm max}]$,
$0 \ll \tilde t_{\rm min}$, $ \tilde t_{\rm max} \ll T$, so that
at any time $t\in [\tilde t_{\rm min}, \tilde t_{\rm
  max}]$, the details of the
distibutions of charges both near the origin
 and outside
  the domain (using axial symmetry) can be essentially neglected,
  keeping only the most slowly decaying contribution. In that
case a
good approximation of $\chi$
is the potential created by a point charge $Q_0$
located at the origin
\begin{equation}\label{eq:potapp}
\chi(\eta,\xi)\approx\frac{Q_0}{\sqrt{\eta^2+\xi^2}} \;,
\end{equation}
with a relation between $Q_0$ and the boundary conditions of the
problem yet to be determined.

The main result of this paper is that the approximation of the charge
distribution by a monopole centered at the origin is precisely the
observed universal behavior expressed by
Eqs.~\eqref{eq:paransatz}-\eqref{eq:scaling}, and in particular the
conditions under which this approximation is valid provide the regime
of validity of the scaling form
Eqs.~\eqref{eq:paransatz}-\eqref{eq:scaling}. Indeed, inserting
Eq.~\eqref{eq:potapp} into Eq.~\eqref{eq:E} and inverting the relation
between $(\xi \smeq v, \eta \smeq \sqrt{|g|m/\mu})$ and $(x,t)$
readily gives
\begin{eqnarray}
m(t,x) &=& \frac{\mu}{9 |g| t^2} \left(\frac{9}{4} |Q_0|^{2/3}\,t^{4/3}
          -x^2\right) \; , \label{eq:parath1} \\
v(t,x) &=& \frac{2}{3} \; \frac{x}{t} \; .
\end{eqnarray}
Equation \eqref{eq:parath1} identifies to \eqref{eq:paransatz} with
$z(t)$ given by Eq.~\eqref{zansatz}. In addition, normalization of
$m(x,t)$ would require that the value of the charge is fixed,
$Q_0 = -2|g|/\mu$. We have now to show that this is actually the case {\em independently of the
  boundary condition} to prove that Eq.~\eqref{eq:parath1} is indeed
the numerically observed scaling form
Eqs.~\eqref{eq:paransatz}-\eqref{eq:scaling}.

For this purpose, we  simply  apply Gauss's law 
\begin{equation} \label{gauss}
Q_0  = \frac{1}{4\pi} \int_{S_{\tilde t}}(\vec E\cdot\vec n)dS\; 
\end{equation}
on a surface ${S_{\tilde t}}$ on which time is constant,
$t(\eta,\xi) = \tilde t$, for some $t\in[0,T]$. Using $x$ and the dumy
angle $\theta$ as coordinates on this surface, the surface element$dS$
and normal vector $\vec n$ read, respectively
$dS =\eta j(x,\tilde t) d\theta dx $ and
$\vec n = (n_\xi,n_\eta,n_\theta) = j(x,\tilde t)^{-1}
\left(\partial_x \xi, -\partial_x \eta, 0 \right) $,
with
$j(x,\tilde t) \equiv \sqrt{(\partial_x\xi)^2+(\partial_x\eta)^2}$.
We get
\begin{eqnarray}
Q_0 & = \frac{1}{4\pi}\int_0^{2\pi}\int_{\mathbb R}\eta\left[2(x-\xi
     \tilde t)\partial_x\eta-\eta \tilde t\partial_x \xi\right]d\theta
     dx \nonumber \\
& = \frac{1}{2}\int_{\mathbb R}\left[-\tilde t\partial_x(\eta^2\xi)+2x\eta\partial_x\eta\right]dx
\; . \label{Q}
\end{eqnarray}
In the absence of mass flow at infinity, the first, time dependent,
term integrates to zero. This  appears here as a consequence of
  the fact that for all $\tilde t\in[0,T]$, the total charge
enclosed in $S_{\tilde t}$ is the
same and $Q_0$ is by construction {\em a constant of
  motion}. Integrating by part the second term
and recalling that $\eta=2\sqrt{{|g|m}/{\mu}}$, Eq.\eqref{Q} yields
$Q_0 = \frac{2g}{\mu} \int_{\mathbb R} m dx$
  \sout{$ Q_0 = \frac{2g}{\mu}$} , which, because of the normalization
condition $\int_{\mathbb R} m dx =1$, is
    the required result.

  To conclude, we have shown that, using the potential representation
  of the [one dimensional potential-free] Mean Field Game problem we
  consider, the remarkable, and a priori quite puzzling {\em universal
    scaling form} which shows up in numerical simulations can be
  derived in a very natural way by approximating the charge
  distribution creating the electrostatic potential $\chi$
  to a simple monopole.  The condition that the
  observation point is sufficiently far from both the charges near the
  origin and the one near infinity underlying this approximation is
  equivalent to considering times far from both $t=0$ and $t=T$, which
  is of course only possible in the limit of very long
  optimization time.  Furthermore,
  the electrostatic picture would allow for a more
  precise quantification of the conditions of validity of the
  approximation, using estimates on boundary conditions.  We have
  thus been allowed to extend in some sense the notion of ergodic
  state to a situation where a genuine ergodic state cannot exist.

At a more general level, the potential representation of our Mean
Field Games underlies the {\em integrability} of the hydrodynamical
equations~\eqref{hydroMFG}. Beyond the simple monopolar
approximation Eq.~\eqref{eq:potapp}, we can construct a complete
multipolar expansion for the potential $\chi$, and it can be seen
 that each ``charge'' in this expansion relates to a
conserved quantity of the dynamics.  The mapping between the boundary
conditions and these charges being thus equivalent to the mapping
between boundary conditions and constants of  motion.  These
considerations emphasize the fact that the deep reason behind the
scaling law characterizing the 1d potential free MFG that we consider
in this paper is actually their integrable character.  A more in depth
discussion of this question will appear in a subsequent publication.

\bibliography{Biblio-2019-07-03,general_ref}

\begin{thebibliography}{13}%
\makeatletter
\providecommand \@ifxundefined [1]{%
 \@ifx{#1\undefined}
}%
\providecommand \@ifnum [1]{%
 \ifnum #1\expandafter \@firstoftwo
 \else \expandafter \@secondoftwo
 \fi
}%
\providecommand \@ifx [1]{%
 \ifx #1\expandafter \@firstoftwo
 \else \expandafter \@secondoftwo
 \fi
}%
\providecommand \natexlab [1]{#1}%
\providecommand \enquote  [1]{``#1''}%
\providecommand \bibnamefont  [1]{#1}%
\providecommand \bibfnamefont [1]{#1}%
\providecommand \citenamefont [1]{#1}%
\providecommand \href@noop [0]{\@secondoftwo}%
\providecommand \href [0]{\begingroup \@sanitize@url \@href}%
\providecommand \@href[1]{\@@startlink{#1}\@@href}%
\providecommand \@@href[1]{\endgroup#1\@@endlink}%
\providecommand \@sanitize@url [0]{\catcode `\\12\catcode `\$12\catcode
  `\&12\catcode `\#12\catcode `\^12\catcode `\_12\catcode `\%12\relax}%
\providecommand \@@startlink[1]{}%
\providecommand \@@endlink[0]{}%
\providecommand \url  [0]{\begingroup\@sanitize@url \@url }%
\providecommand \@url [1]{\endgroup\@href {#1}{\urlprefix }}%
\providecommand \urlprefix  [0]{URL }%
\providecommand \Eprint [0]{\href }%
\providecommand \doibase [0]{http://dx.doi.org/}%
\providecommand \selectlanguage [0]{\@gobble}%
\providecommand \bibinfo  [0]{\@secondoftwo}%
\providecommand \bibfield  [0]{\@secondoftwo}%
\providecommand \translation [1]{[#1]}%
\providecommand \BibitemOpen [0]{}%
\providecommand \bibitemStop [0]{}%
\providecommand \bibitemNoStop [0]{.\EOS\space}%
\providecommand \EOS [0]{\spacefactor3000\relax}%
\providecommand \BibitemShut  [1]{\csname bibitem#1\endcsname}%
\let\auto@bib@innerbib\@empty
\bibitem [{\citenamefont {Lasry}\ and\ \citenamefont
  {Lions}(2006)}]{LasryLions2006-1}%
  \BibitemOpen
  \bibfield  {author} {\bibinfo {author} {\bibfnamefont {J.-M.}\ \bibnamefont
  {Lasry}}\ and\ \bibinfo {author} {\bibfnamefont {P.-L.}\ \bibnamefont
  {Lions}},\ }\href {\doibase http://dx.doi.org/10.1016/j.crma.2006.09.019}
  {\bibfield  {journal} {\bibinfo  {journal} {C. R. Acad. Sci. Paris, Ser. I}\
  }\textbf {\bibinfo {volume} {343}},\ \bibinfo {pages} {619 } (\bibinfo {year}
  {2006})}\BibitemShut {NoStop}%
\bibitem [{\citenamefont {Lachapelle}\ \emph {et~al.}(2015)\citenamefont
  {Lachapelle}, \citenamefont {Lasry}, \citenamefont {Lehalle},\ and\
  \citenamefont {Lions}}]{Lachapelle2015}%
  \BibitemOpen
  \bibfield  {author} {\bibinfo {author} {\bibfnamefont {A.}~\bibnamefont
  {Lachapelle}}, \bibinfo {author} {\bibfnamefont {J.-M.}\ \bibnamefont
  {Lasry}}, \bibinfo {author} {\bibfnamefont {C.-A.}\ \bibnamefont {Lehalle}},
  \ and\ \bibinfo {author} {\bibfnamefont {P.-L.}\ \bibnamefont {Lions}},\
  }\href@noop {} {\bibfield  {journal} {\bibinfo  {journal} {arXiv:1305.6323}\
  } (\bibinfo {year} {2015})}\BibitemShut {NoStop}%
\bibitem [{\citenamefont {Lasry}\ and\ \citenamefont
  {Lions}(2007)}]{lasry_mean_2007}%
  \BibitemOpen
  \bibfield  {author} {\bibinfo {author} {\bibfnamefont {J.-M.}\ \bibnamefont
  {Lasry}}\ and\ \bibinfo {author} {\bibfnamefont {P.-L.}\ \bibnamefont
  {Lions}},\ }\href {\doibase 10.1007/s11537-007-0657-8} {\bibfield  {journal}
  {\bibinfo  {journal} {Japanese Journal of Mathematics}\ }\textbf {\bibinfo
  {volume} {2}},\ \bibinfo {pages} {229} (\bibinfo {year} {2007})}\BibitemShut
  {NoStop}%
\bibitem [{\citenamefont {Gu{\'e}ant}\ \emph {et~al.}(2011)\citenamefont
  {Gu{\'e}ant}, \citenamefont {Lasry},\ and\ \citenamefont
  {Lions}}]{GueantLasryLions2010}%
  \BibitemOpen
  \bibfield  {author} {\bibinfo {author} {\bibfnamefont {O.}~\bibnamefont
  {Gu{\'e}ant}}, \bibinfo {author} {\bibfnamefont {J.-M.}\ \bibnamefont
  {Lasry}}, \ and\ \bibinfo {author} {\bibfnamefont {P.-L.}\ \bibnamefont
  {Lions}},\ }\enquote {\bibinfo {title} {Mean field games and applications},}\
  in\ \href {\doibase 10.1007/978-3-642-14660-2_3} {\emph {\bibinfo {booktitle}
  {Paris-Princeton Lectures on Mathematical Finance 2010}}}\ (\bibinfo
  {publisher} {Springer},\ \bibinfo {address} {Heidelberg},\ \bibinfo {year}
  {2011})\ pp.\ \bibinfo {pages} {205--266}\BibitemShut {NoStop}%
\bibitem [{\citenamefont {Achdou}\ \emph {et~al.}(2014)\citenamefont {Achdou},
  \citenamefont {Buera}, \citenamefont {Lasry}, \citenamefont {P.-L.},\ and\
  \citenamefont {Moll}}]{Achdou2014}%
  \BibitemOpen
  \bibfield  {author} {\bibinfo {author} {\bibfnamefont {Y.}~\bibnamefont
  {Achdou}}, \bibinfo {author} {\bibfnamefont {F.~J.}\ \bibnamefont {Buera}},
  \bibinfo {author} {\bibfnamefont {J.-M.}\ \bibnamefont {Lasry}}, \bibinfo
  {author} {\bibfnamefont {L.}~\bibnamefont {P.-L.}}, \ and\ \bibinfo {author}
  {\bibfnamefont {B.}~\bibnamefont {Moll}},\ }\href
  {http://dx.doi.org/10.1098/rsta.2013.0397} {\bibfield  {journal} {\bibinfo
  {journal} {Phil. Trans. R. Soc. A}\ }\textbf {\bibinfo {volume} {372}},\
  \bibinfo {pages} {20130397} (\bibinfo {year} {2014})}\BibitemShut {NoStop}%
\bibitem [{\citenamefont {M\'eriauxi}\ \emph {et~al.}(2012)\citenamefont
  {M\'eriauxi}, \citenamefont {Varma},\ and\ \citenamefont
  {Lasaulce}}]{WirelessNetwork}%
  \BibitemOpen
  \bibfield  {author} {\bibinfo {author} {\bibfnamefont {F.}~\bibnamefont
  {M\'eriauxi}}, \bibinfo {author} {\bibfnamefont {V.}~\bibnamefont {Varma}}, \
  and\ \bibinfo {author} {\bibfnamefont {S.}~\bibnamefont {Lasaulce}},\ }in\
  \href {\doibase 10.1109/ACSSC.2012.6489095} {\emph {\bibinfo {booktitle}
  {2012 Conference Record of the Forty Sixth Asilomar Conference on Signals,
  Systems and Computers (ASILOMAR)}}}\ (\bibinfo {year} {2012})\ pp.\ \bibinfo
  {pages} {671--675}\BibitemShut {NoStop}%
\bibitem [{\citenamefont {Kizilkale}\ and\ \citenamefont
  {Malhamé}(2016)}]{KizilkaleMalhame2015}%
  \BibitemOpen
  \bibfield  {author} {\bibinfo {author} {\bibfnamefont {A.}~\bibnamefont
  {Kizilkale}}\ and\ \bibinfo {author} {\bibfnamefont {R.}~\bibnamefont
  {Malhamé}},\ }in\ \href {\doibase
  https://doi.org/10.1016/B978-0-12-805246-4.00020-3} {\emph {\bibinfo
  {booktitle} {Control of Complex Systems}}},\ \bibinfo {editor} {edited by\
  \bibinfo {editor} {\bibfnamefont {K.~G.}\ \bibnamefont {Vamvoudakis}}\ and\
  \bibinfo {editor} {\bibfnamefont {S.}~\bibnamefont {Jagannathan}}}\ (\bibinfo
   {publisher} {Butterworth-Heinemann},\ \bibinfo {year} {2016})\ pp.\ \bibinfo
  {pages} {559 -- 584}\BibitemShut {NoStop}%
\bibitem [{\citenamefont {Bertsekas}(2017)}]{bertsekas2012}%
  \BibitemOpen
  \bibfield  {author} {\bibinfo {author} {\bibfnamefont {D.}~\bibnamefont
  {Bertsekas}},\ }\href@noop {} {\emph {\bibinfo {title} {Dynamic Programming
  and Optimal Control}}}\ (\bibinfo  {publisher} {Athena Scientific},\ \bibinfo
  {address} {Nashua},\ \bibinfo {year} {2017})\BibitemShut {NoStop}%
\bibitem [{\citenamefont {Ullmo}\ \emph {et~al.}(2019)\citenamefont {Ullmo},
  \citenamefont {Swiecicki},\ and\ \citenamefont {Gobron}}]{ULLMO20191}%
  \BibitemOpen
  \bibfield  {author} {\bibinfo {author} {\bibfnamefont {D.}~\bibnamefont
  {Ullmo}}, \bibinfo {author} {\bibfnamefont {I.}~\bibnamefont {Swiecicki}}, \
  and\ \bibinfo {author} {\bibfnamefont {T.}~\bibnamefont {Gobron}},\ }\href
  {\doibase https://doi.org/10.1016/j.physrep.2019.01.001} {\bibfield
  {journal} {\bibinfo  {journal} {Physics Reports}\ }\textbf {\bibinfo {volume}
  {799}},\ \bibinfo {pages} {1 } (\bibinfo {year} {2019})}\BibitemShut
  {NoStop}%
\bibitem [{\citenamefont {Cardaliaguet}\ \emph {et~al.}(2013)\citenamefont
  {Cardaliaguet}, \citenamefont {Lasry}, \citenamefont {Lions},\ and\
  \citenamefont {Porretta}}]{Cardaliaguet2013}%
  \BibitemOpen
  \bibfield  {author} {\bibinfo {author} {\bibfnamefont {P.}~\bibnamefont
  {Cardaliaguet}}, \bibinfo {author} {\bibfnamefont {J.-M.}\ \bibnamefont
  {Lasry}}, \bibinfo {author} {\bibfnamefont {P.-L.}\ \bibnamefont {Lions}}, \
  and\ \bibinfo {author} {\bibfnamefont {A.}~\bibnamefont {Porretta}},\ }\href
  {\doibase 10.1137/120904184} {\bibfield  {journal} {\bibinfo  {journal} {SIAM
  J. Control Optim.}\ }\textbf {\bibinfo {volume} {51}},\ \bibinfo {pages}
  {3558} (\bibinfo {year} {2013})},\ \Eprint
  {http://arxiv.org/abs/http://dx.doi.org/10.1137/120904184}
  {http://dx.doi.org/10.1137/120904184} \BibitemShut {NoStop}%
\bibitem [{\citenamefont {Pethick}\ and\ \citenamefont
  {Smith}(2008)}]{pethick_smith_2008}%
  \BibitemOpen
  \bibfield  {author} {\bibinfo {author} {\bibfnamefont {C.~J.}\ \bibnamefont
  {Pethick}}\ and\ \bibinfo {author} {\bibfnamefont {H.}~\bibnamefont
  {Smith}},\ }\href {\doibase 10.1017/CBO9780511802850} {\emph {\bibinfo
  {title} {Bose-Einstein Condensation in Dilute Gases}}},\ \bibinfo {edition}
  {2nd}\ ed.\ (\bibinfo  {publisher} {Cambridge University Press},\ \bibinfo
  {year} {2008})\BibitemShut {NoStop}%
\bibitem [{\citenamefont {Dalfovo}\ \emph {et~al.}(1999)\citenamefont
  {Dalfovo}, \citenamefont {Giorgini}, \citenamefont {Pitaevskii},\ and\
  \citenamefont {Stringari}}]{Dalfovo1999Theory}%
  \BibitemOpen
  \bibfield  {author} {\bibinfo {author} {\bibfnamefont {F.}~\bibnamefont
  {Dalfovo}}, \bibinfo {author} {\bibfnamefont {S.}~\bibnamefont {Giorgini}},
  \bibinfo {author} {\bibfnamefont {L.~P.}\ \bibnamefont {Pitaevskii}}, \ and\
  \bibinfo {author} {\bibfnamefont {S.}~\bibnamefont {Stringari}},\ }\href
  {\doibase 10.1103/revmodphys.71.463} {\bibfield  {journal} {\bibinfo
  {journal} {Reviews of Modern Physics}\ }\textbf {\bibinfo {volume} {71}},\
  \bibinfo {pages} {463} (\bibinfo {year} {1999})}\BibitemShut {NoStop}%
\bibitem [{\citenamefont {Kamchatnov}(2000)}]{kamchatnov2000nonlinear}%
  \BibitemOpen
  \bibfield  {author} {\bibinfo {author} {\bibfnamefont {A.}~\bibnamefont
  {Kamchatnov}},\ }\href {https://books.google.fr/books?id=yM99QgAACAAJ} {\emph
  {\bibinfo {title} {Nonlinear Periodic Waves and Their Modulations: An
  Introductory Course}}}\ (\bibinfo  {publisher} {World Scientific},\ \bibinfo
  {year} {2000})\BibitemShut {NoStop}%
\end{thebibliography}%
\end{document}